\documentclass[sigconf, screen]{acmart}

\usepackage{multirow}
\usepackage{booktabs}
\usepackage{algorithm}
\usepackage{algorithmic}
\usepackage{xcolor}
\usepackage{listings}
\usepackage[capitalize]{cleveref}
\usepackage{amsmath}

\crefname{section}{Sec.}{Secs.}
\Crefname{section}{Section}{Sections}
\Crefname{table}{Table}{Tables}
\Crefname{algorithm}{Algorithm}{Algorithms}
\crefname{equation}{Equation}{Equations}

\crefname{table}{Tab.}{Tabs.}
\crefname{algorithm}{Alg.}{Algs.}
\crefname{lemma}{Lem.}{Lems.}
\crefname{theorem}{Thm.}{Thms.}
\crefname{definition}{Def.}{Defs.}
\crefname{equation}{Eq.}{Eqs.}

\renewcommand\footnotetextcopyrightpermission[1]{}
\DeclareMathOperator*{\argmin}{arg\,min}
\AtBeginDocument{%
  }

\setcopyright{acmcopyright}
\copyrightyear{2023}
\acmYear{2023}
\acmDOI{XXXXXXX.XXXXXXX}

\acmConference[]{}{}{}
\acmPrice{15.00}
\acmISBN{978-1-4503-XXXX-X/18/06}

\acmSubmissionID{2224}

\begin{document}

\title [Flexible Differentially Private Vertical Federated Learning with Adaptive Feature Embeddings] {Flexible Differentially Private Vertical Federated Learning \\
with Adaptive Feature Embeddings}

\author{Yuxi Mi}
\email{yxmi20@fudan.edu.cn}
\affiliation{%
  \institution{Fudan University}
  \city{Shanghai}
  \country{China}
}

\author{Hongquan Liu}
\email{hqliu21@m.fudan.edu.cn}
\affiliation{%
  \institution{Fudan University}
  \city{Shanghai}
  \country{China}
}

\author{Yewei Xia}
\email{ywxia21@m.fudan.edu.cn}
\affiliation{%
  \institution{Fudan University}
  \city{Shanghai}
  \country{China}
}

\author{Yiheng Sun}
\email{elisun@tencent.com}
\affiliation{%
  \institution{Tencent}
  \city{Shenzhen}
  \country{China}
}

\author{Jihong Guan}
\email{jhguan@tongji.edu.cn}
\affiliation{%
  \institution{Tongji University}
  \city{Shanghai}
  \country{China}
}

\author{Shuigeng Zhou}
\email{sgzhou@fudan.edu.cn}
\affiliation{%
  \institution{Fudan University}
  \city{Shanghai}
  \country{China}
}

\renewcommand{\shortauthors}{Yuxi Mi et al.}

\begin{abstract}
The emergence of vertical federated learning (VFL) has stimulated concerns about the imperfection in privacy protection, as shared feature embeddings may reveal sensitive information under privacy attacks. This paper studies the delicate equilibrium between data privacy and task utility goals of VFL under differential privacy (DP). To address the generality issue of prior arts, this paper advocates a flexible and generic approach that decouples the two goals and addresses them successively. Specifically, we initially derive a rigorous privacy guarantee by applying norm clipping on shared feature embeddings, which is applicable across various datasets and models. Subsequently, we demonstrate that task utility can be optimized via adaptive adjustments on the scale and distribution of feature embeddings in an accuracy-appreciative way, without compromising established DP mechanisms. We concretize our observation into the proposed VFL-AFE framework, which exhibits effectiveness against privacy attacks and the capacity to retain favorable task utility, as substantiated by extensive experiments.

\end{abstract}




\maketitle

\section{Introduction}
\label{sec:intro}

\begin{figure}[tbp]
  \centering
  \includegraphics[width=\linewidth]{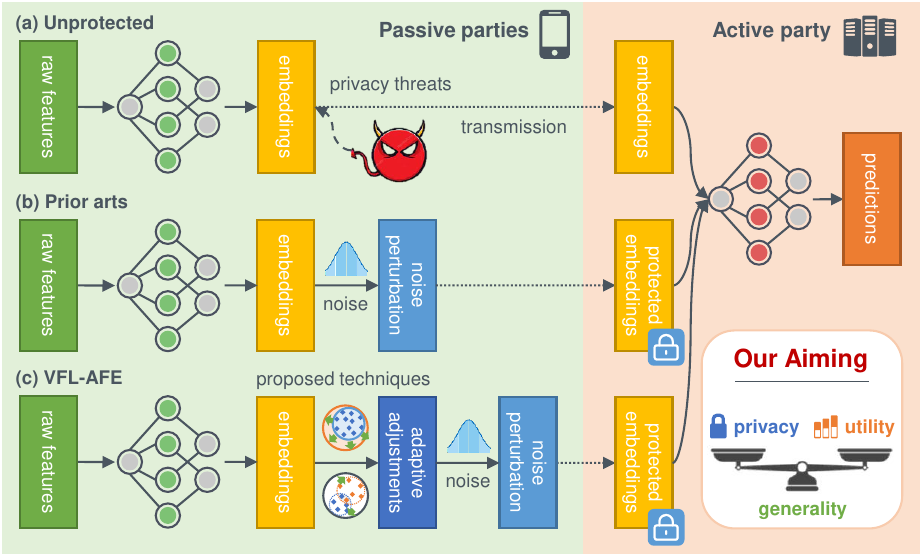}
  \caption{Paradigm comparison among unprotected VFL, prior arts, and our VFL-AFE. (a) Unprotected VFL directly shares feature embeddings, making them prone to privacy threats. (2) Prior arts calculate delicate noise scales to balance privacy and utility, but their generality is limited. (3) Our VFE-AFL adopts a more flexible and generalizable approach to enhance privacy and utility separately.}
  \Description{Comparison of unprotected VFL, prior arts, and our proposed VFL-AFE method in paradigm.}
  \label{fig:paradigm}
  \vspace{-3mm}
\end{figure}

Federated learning (FL)~\cite{DBLP:conf/aistats/McMahanMRHA17} is a rapidly evolving machine learning approach that facilitates collaborative model establishment across multiple parties, each holding a partition of the dataset, by synchronizing local computation results without centralizing the data. FL can be categorized into either horizontal or vertical paradigms, depending on how the data is partitioned in the sample and feature space~\cite{liu2022vertical}. Due to its privacy awareness, FL has gained increasing adoption recently in response to the growing regulatory demands.

This paper investigates the trade-off between data privacy and task utility in feature-partitioned \textit{vertical federated learning} (VFL)~\cite{DBLP:journals/corr/XieCZL22Admm,DBLP:conf/kdd/HuNYZ19,DBLP:journals/corr/HuLKN19Admm,DBLP:journals/corr/WangLHBBL20Hybrid,DBLP:journals/corr/RanbadugeD22DPVFL,DBLP:journals/corr/ChenJSY20VAFL,DBLP:conf/ccs/XuB00JL21,DBLP:journals/corr/CeballosSMSRVR20SplitNN,DBLP:journals/corr/CastigliaWP22Flexible,DBLP:journals/tsp/LiuZKLCHY22,10.1145/3467956,DBLP:journals/corr/HardyHINPST17VFLHE,DBLP:journals/corr/YangFCSY19QuasiNewton}.  In VFL, multiple parties jointly train a model by sharing a common set of data instances, while each party holds partial feature dimensions or labels.  At each communication round, the feature-holding parties (``\textit{passive parties}'') exchange feature embeddings extracted from their private local models, which may be heterogeneous and of varying forms. A label-holding coordinator (``\textit{active party}'') aggregates the embeddings and returns calculated gradients, therefrom the passive parties update their local models. VFL has demonstrated broad potential in applications~\cite{DBLP:journals/tist/YangLCT19,DBLP:journals/corr/HardRMBAEKR18keyboard} where parties possess \textit{different sources, views, and modalities of data} regarding the same subject. For instance, by building joint models on medical visits and prescription records, healthcare institutions could gain a comprehensive understanding of a patient’s health condition.

Despite the privacy-aware designs of VFL, there are still sparkling concerns regarding the imperfections in data protection. Recent studies have shown that shared feature embeddings can undesirably expose sensitive information of passive parties under privacy threats, such as \textit{inversion} and \textit{membership inference} (MI) attacks. Inversion attacks~\cite{DBLP:conf/icde/LuoWXO21,DBLP:conf/acsac/HeZL19,DBLP:journals/popets/JiangZG22a,DBLP:journals/corr/YeJWLL22recon,DBLP:journals/corr/WengZXWJZ20leakage,DBLP:journals/corr/JinCHYC21cafe} enable the recovery of raw features from embeddings (\textit{e.g.} recovering clinical records from diagnoses). On the other hand, membership inference attacks~\cite{DBLP:conf/sp/ShokriSSS17,DBLP:conf/ccs/0001MMBS22,zari2021efficient} allow inferring the presence of certain attributes or subjects in the database (\textit{e.g.} determining whether a person is within the patient list). Therefore, the data privacy of passive parties can still be seriously compromised if no further measures are taken (\cref{fig:paradigm}(a)).

This paper proposes a novel privacy-preserving VFL framework, VFL-AFE, based on differential privacy. Our approach rigorously ensures data privacy while maintaining decent task utility.

Differential privacy (DP)~\cite{DBLP:journals/fttcs/DworkR14,Dwork14Gaussian}  is a computationally efficient privacy protection technique with extensive use in FL. It obfuscates and de-identifies individual instances while retaining the statistical property of the entirety by adding controlled noise~\cite{Near21ProgrammingDP}.  The primary challenge of DP is to effectively balance the competing data privacy and task utility as the introduced noise perturbations would inevitably affect model accuracy.  Most prior arts~\cite{DBLP:journals/corr/HuLKN19Admm,DBLP:journals/corr/XieCZL22Admm,DBLP:journals/corr/WangLHBBL20Hybrid,DBLP:journals/corr/RanbadugeD22DPVFL,DBLP:conf/kdd/HuNYZ19} employ quite inflexible trade-offs: they take into account \textit{dedicated conditions} regarding training data, loss functions or model architectures, to calibrate delicate noise scales. However, their derivations often rely on specific assumptions, such as model convexity and continuity of loss functions. Although these assumptions facilitate tight noise scales, their privacy guarantees may not be readily applicable in more general settings. (\cref{fig:paradigm}(b)). 
To achieve generality, this paper advocates the following takeaway message: \textit{DP and VFL can be combined in a more flexible way}. Specifically, to \textit{decouple privacy and utility into two separate goals} and address them successively.

We first address data privacy. To achieve formal privacy guarantees, DP typically chooses a noise scale proportional to \textit{sensitivity}, a measurement of the maximum disparity among shared outputs. As it is often nontrivial to derive a closed form of sensitivity, prior arts mostly enforce it to a derived threshold by employing norm clipping on \textit{raw features} and/or \textit{model parameters}. Their dedicated derivations often achieve tight noise scales while sacrificing generality. In this paper, we propose a simple yet effective technique to perform norm clipping directly on output \textit{feature embeddings}.  This enables us to omit assumptions such as specific model architectures from our calculations and to establish a privacy guarantee suitable for generic deep neural networks (DNN).

However, as an equilibrium to generality, our calculation could result in a less tight noise scale, which is unfavorable for task utility. We subsequently reconcile the drawback by employing the proposed \textit{adaptive feature embedding}. We start with a key observation regarding DP’s property: informally, local manipulations of feature embeddings before noise perturbation will not impair privacy (by consuming privacy budgets), as long as no additional information is publicly shared and the required noise scale remains unchanged. Hence, we can locally adjust feature embeddings \textit{before} adding noise, in a manner appreciative for accuracy, without compromising established DP mechanisms (\cref{fig:paradigm}(c)). We concretize the observation into two related techniques: (1) We \textit{rescale} the feature embeddings to bridge the gap between actual maximum disparity and estimated sensitivity, allowing full utilization of the noise; (2) We \textit{adjust the distribution} of feature embeddings through weakly-supervised contrastive learning to enhance their inter-class distinguishability, which is beneficial for classification tasks.

In summary, our paper presents three-fold contributions:
\begin{itemize}
    \item We propose a novel differentially private VFL that provides generic privacy guarantees by norm clipping on passive parties’ shared feature embeddings.
    \item We introduce adaptive feature embeddings to enhance task utility, which, to the best of our knowledge, is the first in VFL literature. Specifically, we propose rescaling and adjusting the distribution of feature embeddings.
    \item We present VFL-AFE to concretize our findings. Experiments show VFL-AFE enhances VFL’s task utility while maintaining data privacy, in a generally applicable way. 
\end{itemize}

\section{Related Work}
\label{sec:related-work}

\noindent \textbf{Vertical federated learning (VFL).} VFL~\cite{DBLP:journals/corr/XieCZL22Admm,DBLP:conf/kdd/HuNYZ19,DBLP:journals/corr/HuLKN19Admm,DBLP:journals/corr/WangLHBBL20Hybrid,DBLP:journals/corr/RanbadugeD22DPVFL,DBLP:journals/corr/ChenJSY20VAFL,DBLP:conf/ccs/XuB00JL21,DBLP:journals/corr/CeballosSMSRVR20SplitNN,DBLP:journals/corr/CastigliaWP22Flexible,DBLP:journals/tsp/LiuZKLCHY22,10.1145/3467956,DBLP:journals/corr/HardyHINPST17VFLHE,DBLP:journals/corr/YangFCSY19QuasiNewton} is feature-partitioned federated learning~\cite{DBLP:conf/aistats/McMahanMRHA17,liu2022vertical}, where parties share common sample space while each holding different feature dimensions. Pioneering studies of VFL are mostly based on simple machine learning models such as trees~\cite{DBLP:journals/pvldb/WuCXCO20} and linear classifiers~\cite{DBLP:journals/corr/HuLKN19Admm,DBLP:journals/corr/WangLHBBL20Hybrid}. In light of split learning~\cite{DBLP:journals/corr/CeballosSMSRVR20SplitNN}, recent advances extend the capability of VFL into generic DNNs. 

\noindent \textbf{Threats to data privacy.} By imposing attacks, sensitive information of the passive parties could still be exposed through shared feature embeddings. Specifically, inversion attacks~\cite{DBLP:conf/icde/LuoWXO21,DBLP:conf/acsac/HeZL19,DBLP:journals/popets/JiangZG22a,DBLP:journals/corr/YeJWLL22recon,DBLP:journals/corr/WengZXWJZ20leakage,DBLP:journals/corr/JinCHYC21cafe} enable recovering of raw features from shared embeddings. Membership inference (MI) attacks~\cite{DBLP:conf/sp/ShokriSSS17,DBLP:conf/ccs/0001MMBS22,zari2021efficient} reveal the presence of specific data instances in training datasets.

\noindent \textbf{Privacy-aware VFL.} Recent studies witness significant advances regarding data privacy in VFL. We broadly divide their means into three categories: (1) Hardware-based methods exploit trusted execution environments~\cite{10.1145/3467956}. (2) Cryptographic methods protect communication with crypto-primitives such as secure multi-party computation~\cite{10.1145/3467956}, homomorphic encryption~\cite{DBLP:journals/corr/HardyHINPST17VFLHE,DBLP:journals/corr/YangFCSY19QuasiNewton}, and functional encryption~\cite{DBLP:conf/ccs/XuB00JL21}. Their bottlenecks are typically high time and computational costs. (3) Perturbation-based methods that modify or regenerate communicated messages~\cite{liu2022vertical}, which are often concretized by differential privacy (DP)~\cite{DBLP:journals/corr/HuLKN19Admm,DBLP:journals/corr/XieCZL22Admm,DBLP:journals/corr/WangLHBBL20Hybrid,DBLP:journals/corr/RanbadugeD22DPVFL,DBLP:conf/kdd/HuNYZ19} mechanisms. Prior arts mostly add Gaussian noise on the raw features or model parameters, where conditions are applied to their derivations, limiting their generic use.~\cite{DBLP:journals/corr/XieCZL22Admm} is most related to this work as they also propose noise on embeddings. However, they address utility by relaxed forms of DP notions, while we by the novel proposed adaptive feature embeddings.

\section{Methodology}
\label{sec:method}

\begin{figure*}[tbp]
  \centering
  \includegraphics[width=\linewidth]{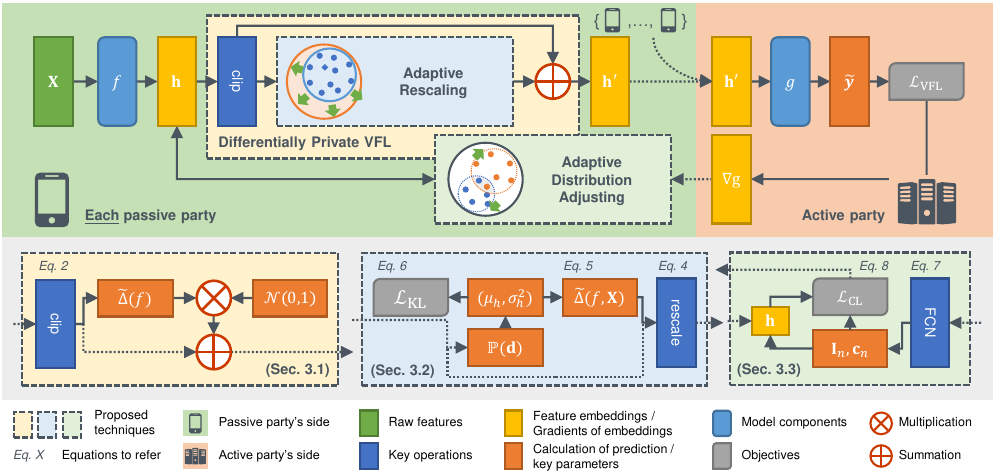}
  \caption{The pipeline of VFL-AFE, which addresses privacy and utility separately. Generic noise perturbation is added to feature embeddings to achieve differential privacy. To enhance task utility, adaptive rescaling of the feature embeddings reduces excessive noise, while adjusting their distributions promotes inter-class discrepancy, thereby improving downstream tasks.}
  \Description{Pipeline of VFL-AFE.}
  \label{fig:pipeline}
  \vspace{-1mm}
\end{figure*}

We here describe the proposed VFL framework with \textit{A}daptive \textit{F}eature \textit{E}mbeddings, referred to as VFL-AFE. In \cref{subsec:dp-vfl}, we first address data privacy by introducing a privacy-preserving VFL that adds calculated noise to the passive parties' output feature embeddings. The method provides formal DP privacy guarantees on generic DNNs. We further dig into the task utility issue under noise perturbation, by showing that the model accuracy can be flexibly improved from adaptive adjustments on the scale and distribution of feature embeddings, as respectively discussed in~\cref{subsec:rescale,subsec:distribution}. \Cref{fig:pipeline} illustrates the pipeline of VFL-AFE.

\begin{algorithm}[tbp]
\begin{algorithmic} [1]
    \REQUIRE Number of parties $M$, number of samples $N$, training data $\mathbf{D} = (\mathbf{X}^1, \mathbf{X}^2, \dots, \mathbf{X}^{M-1}, \mathbf{y})$, batch size $n$, feature extractor models $\{f(\cdot;\theta^i)\}_{i=1}^{M-1}$, head model $g(\cdot;\theta^M)$, \colorbox{yellow!20}{privacy budget and loss $(\epsilon,\delta)$, clipping threshold $\{t^i\}_{i=1}^{M-1}$,} \colorbox{green!20}{number of classes $C$, filtering threshold $c$}.
    \FOR{each communication round}
        \STATE \textbf{conduct} entity alignment, sample $n$ indices from $N$
        \FOR{passive party $i\in[M-1]$}
            \STATE \textbf{generates} local training mini-batch $\mathbf{X}_n^i$
            \STATE \textbf{computes} embeddings $\mathbf{h}_n^i\triangleq \{h_j^i\}_{j\in[n]} \leftarrow f^i(\mathbf{X}_n^i,\theta^i)$
            \STATE \colorbox{yellow!20}{\textbf{clips} norm $\mathbf{h}_n^i\leftarrow {h_j^i}/\max{(1,\frac{||h_j^i||}{t^i})}, \forall j \in [n]$}
            \STATE \colorbox{blue!20}{\textbf{estimates} local sensitivity $\tilde{\Delta}(f^i,\mathbf{X}^i)\leftarrow \mathbf{h}_n^i$}
            \STATE \colorbox{blue!20}{\textbf{rescales} embeddings ${h_j^i} \leftarrow  {h_j^i} / \frac{\tilde{\Delta}(f^i,\mathbf{X}^i)}{2t^i}, \forall j \in [n]$}
            \STATE \colorbox{yellow!20}{\textbf{adds} DP noise ${h'}_j^i \leftarrow  {h_j^i} + \mathcal{N}(0,4\sigma^2 {t^i}^2), \forall j \in [n]$}
            \STATE \textbf{shares} $\mathbf{h}_n^i$ with the active party $M$
        \ENDFOR
        \STATE $M$ \textbf{concatenates} $\mathbf{h}_n\leftarrow\{\mathbf{h}_n^i\}_{i\in[M-1]}$
        \STATE $M$ \textbf{optimizes} $\mathcal{L}(g(\mathbf{h}_n;\theta^M);\mathbf{y}_n)$ and obtains $\nabla g_n^i$
        \STATE $M$ \textbf{exchanges} $\nabla g_n^i$ with passive party $i,\forall i\in [M-1]$
        \FOR{passive party $i\in[M-1]$}
            \STATE \colorbox{green!20}{\textbf{performs} fuzzy clustering $\{\mathbf{I}_n^i,\mathbf{c}_n^i\} \leftarrow \operatorname{FCM}(\nabla g_n^i, C)$}
            \STATE \textbf{calculates} locally \colorbox{blue!20}{$\mathcal{L}_{KL}^i\leftarrow \mathbf{h}_n^i$}\colorbox{green!20}{, $\mathcal{L}_{CL}^i\leftarrow \{\mathbf{h}_n^i,\mathbf{I}_n^i,\mathbf{c}_n^i\}$}
            \STATE \textbf{updates} $\theta^i$ via SGD with $\nabla g_n^i$, $r(\theta^i)$\colorbox{blue!20}{, $\mathcal{L}_{KL}^i$}\colorbox{green!20}{, $\mathcal{L}_{CL}^i$}
        \ENDFOR
    \ENDFOR

\end{algorithmic}

\caption{The proposed VFL-AFE framework}
\label{alg:vfl-afe}
\end{algorithm}

\subsection{Differentially Private VFL}
\label{subsec:dp-vfl}

We start by formulating the VFL framework. VFL is designed for the distributed training of models among a set of $M$ parties, who hold the same or similar data samples yet are partitioned by feature dimensions. The training is initiated and supervised by the sole party who owns the labels, referred to as the \textit{active party}. We denote the $M$-th party as the active party, \textit{wlog.}, and the remaining ($M$-1) feature-holding parties as the \textit{passive parties}.

We denote the VFL dataset with $N$ training samples as $\mathbf{D} = (\mathbf{X}, \mathbf{y}) = (\mathbf{X}^1, \mathbf{X}^2, \dots, \mathbf{X}^{M-1}, \mathbf{y})$, where $\mathbf{X}^i \triangleq \{x_{j}^{i}\}_{j=1}^{N}$ is the local feature vector set owned by the $i$-th passive party and $\mathbf{y} \triangleq \{y_j\}_{j=1}^{N}$ is the label set. We assume $(\mathbf{X}, \mathbf{y})$ are aligned by data sample, \textit{i.e.}, $(x_j^1,x_j^2, \dots, x_j^{M-1},y_j)$ are partitioned from the same $(x_j,y_j)$, $\forall j$. This can be achieved by private set intersection (PSI)~\cite{LuN20PSI}.


To prevent centralizing the data, each passive party $i$ locally learns a feature extractor model $f^i(\cdot)$ parameterized by $\theta^i$ that maps its raw features $\mathbf{X}^i$ into low-dimensional feature embeddings $\mathbf{h}^i \triangleq \{h_j^i\}_{j=1}^{N} = f^i(\mathbf{X}^i; \theta^i)$, then shares $\mathbf{h}^i$ with the active party. The active party aggregates all $\{\mathbf{h}^i\}_{i=1}^{M-1}$ by concatenating them to train a head model $g(\cdot)$ parameterized by $\theta^M$ which produces final predictions. All parties aim to collaboratively solve the objective:

\begin{equation}
\label{eq:objective}
    {\argmin_{\{\theta^i\}_{i=1}^{M}}} {\frac{1}{N}\sum_{j=1}^{N} {\mathcal{L}(g(h_j^1,h_j^2,\dots,h_j^{M-1};\theta^M);y_j)}} +
    \lambda \sum_{i=1}^{M} r(\theta^i),
\end{equation}

\noindent where $\mathcal{L}$ is a generic supervised loss function (\textit{e.g.} a cross-entropy loss with softmax activation) and $r(\cdot)$ is the party-wise regularization term together weighted by $\lambda$. To update the model, the active party calculates and exchanges the gradients $\nabla^i{g}$ with respect to each $\mathbf{h}^i$ and the passive parties update their extractors therefrom. \Cref{alg:vfl-afe} presents the process of our VFL-AFE method, where the colored texts highlight our key techniques: \colorbox{yellow!20}{noise perturbation}, \colorbox{blue!20}{rescaling}, and \colorbox{green!20}{distribution adjusting}, which will be discussed in detail later.

To mitigate the risk of potential data leakage, we let the passive parties introduce randomized noise during their computation of feature embeddings, which obfuscates the fine-grained details of individual data instances from the observation of the active party and any third-party adversaries. The noise is quantitatively measured by \textit{differential privacy}. We briefly revisit its key notions.

\begin{definition} [Differential Privacy~\cite{DBLP:journals/fttcs/DworkR14}]
\label{def:dp}
Denote $\mathbf{D},\mathbf{D}' \in \mathcal{D}$ over domain $\mathcal{D}$ that differ by exactly one data instance as \textit{neighboring} datasets. A randomized algorithm $\mathcal{A}: \mathcal{D} \rightarrow \mathcal{R}$ with range $\mathcal{R}$ satisfies $(\epsilon,\delta)$-differential privacy if for any two neighboring $\mathbf{D},\mathbf{D}'$ and any set of outputs $\mathcal{O} \in \mathcal{R}$, the following holds:

$$\mathbb{P}[\mathcal{A}(\mathbf{D})\in \mathcal{O}] \leq \exp{(\epsilon)} \mathbb{P}[\mathcal{A}(\mathbf{D}')\in \mathcal{O}] + \delta.$$

\end{definition}

\noindent $\mathcal{A}$ satisfying DP is called a \textit{mechanism}. The pair $(\epsilon,\delta)$ is referred to as \textit{privacy budget} and \textit{loss}, where smaller $\epsilon,\delta$ informally indicates a better level of protection and lower failure probability of $\mathcal{A}$, respectively. Specifically, the scale of noise required to ensure differential privacy of $\mathcal{A}$ depends on the \textit{sensitivity}, which describes the maximum disparity of $\mathcal{A}$ between $\mathbf{D},\mathbf{D}'$.

\begin{definition} [Sensitivity~\cite{DBLP:journals/fttcs/DworkR14}]
\label{def:sensitivity}
    The sensitivity of a function $f: \mathbf{D}\rightarrow \mathbb{R}^l$ under any neighboring $\mathbf{D},\mathbf{D}'$ is defined as:
    $$\Delta(f)=\max_{\mathbf{D},\mathbf{D}'} ||f(\mathbf{D})-f(\mathbf{D}')||.$$
 
\end{definition}

\noindent Here, $||\cdot||$ denotes the distance metric required by a particular mechanism. We adopt $l_2$ norm for the \textit{Gaussian mechanism}~\cite{Dwork14Gaussian}, which associates the quantity of noise with the desired level of privacy.

\begin{lemma} [Gaussian Mechanism~\cite{Dwork14Gaussian}]
\label{lem:gaussian}
Let $f: \mathbf{D}\rightarrow \mathbb{R}^l$ be an arbitrary function. For any $\epsilon \in (0,1)$, choose $c^2 > 2\log(\frac{1.25}{\delta})$. Then, $f+\mathcal{N}(0,(\sigma\Delta(f))^2)$ with $\sigma \geq \frac{c}{\epsilon}$ satisfies $(\epsilon,\delta)$-differential privacy.
\end{lemma}

According to the above notions, given $(\epsilon,\delta)$, a noise scale can be calibrated proportionally to sensitivity $\Delta(f)$ based on \cref{lem:gaussian}.  Then, by releasing $\mathbf{h}^i$ with respective noise $ \mathcal{N}(0,(\sigma\Delta(f))^2)$ , the passive parties can safeguard the privacy of $\mathbf{h}^i$ through formal DP guarantees.  The final piece of puzzle unsolved here is sensitivity. In practice, an estimation $\tilde{\Delta}(f)$ of $\Delta(f)$ is commonly derived (as it's often infeasible to calculate the exact $\Delta(f)$), which choice involves a trade-off between privacy and accuracy: Privacy would be compromised if one wrongfully chooses $\tilde{\Delta}(f)<\Delta(f)$ while choosing $\tilde{\Delta}(f)\gg \Delta(f)$ would introduce excessive noise that impairs task utility. Generally, DP demands \textit{a tight $\tilde{\Delta}(f)$ that introduces minimal noise while satisfying the desired privacy level $(\epsilon,\delta)$}.

To determine an appropriate $\tilde{\Delta}(f)$, \textit{norm clipping} is commonly applied to enforce the range of local outputs to a specified threshold. In VFL, most prior arts~\cite{DBLP:journals/corr/HuLKN19Admm,DBLP:journals/corr/WangLHBBL20Hybrid,DBLP:journals/corr/RanbadugeD22DPVFL,DBLP:conf/kdd/HuNYZ19} employ norm clipping on \textit{raw features} $\mathbf{X}^i$ and/or \textit{model parameters} $\theta^i$ to indirectly constrain the range of feature embeddings $\mathbf{h}^i$ (as $\mathbf{h}^i = f^i(\mathbf{X}^i; \theta^i)$), and derive $\tilde{\Delta}(f)$ therefrom. Their derived $\tilde{\Delta}(f)$ are often tight however at the cost of limited generality: They either dependent on specific model structures (\textit{e.g.}, trees or linear classifiers) or rely on certain theoretical assumptions (\textit{e.g.}, model convexity and the Lipschitz continuity of the loss function), which may not hold in many cases. 

We propose a simple yet effective \colorbox{yellow!20}{technique} regarding the issue of generality. Specifically, we perform norm clipping directly on \textit{feature embeddings} $\mathbf{h}^i$. For a given party-wise \textit{clipping threshold} $t^i$ of passive party $i$, we divide the norm of its feature embeddings $\{h_j^i\}_{j=1}^N$ by $\max{(1,{||h_j^i||}/{t^i})}$.  Note this will enforce the maximum disparity of $f^i$ (thus, the actual sensitivity $\Delta^i(f^i)$) to be no greater than $2t^i$, according to~\cref{def:sensitivity} (please further refer to the proof of~\cref{th:vfl-dp}).
Hence, we can conveniently choose the estimation as $\tilde{\Delta}^{i}(f^i)=2t^i$ for passive party $i$. We let each passive party produce noisy feature embeddings $\mathbf{h}'^i$, as: 


\begin{equation}
    \label{eq:norm-clip}
    {h'}_j^i = {h_j^i}/\max{(1,\frac{||h_j^i||}{t^i})} +\mathcal{N}(0,4\sigma^2 {t^i}^2), \forall j \in [N],
\end{equation}

\noindent and share $\mathbf{h}'^i$ with the active party in replacement of the unprotected raw $\mathbf{h}^i$. We argue after applying \cref{eq:norm-clip}, our VFL framework achieves the following privacy guarantees:

\begin{theorem}
\label{th:vfl-dp}
The VFL framework specified in~\cref{alg:vfl-afe} and modified by \cref{eq:norm-clip} is $(\epsilon,\delta)$-differentially private.
\end{theorem}

The proof is deferred to supplementary materials. 
To briefly summarize, we address data privacy by establishing a privacy-preserving VFL that protects the passive parties' shared outputs under formal DP guarantees. For the convenience of  discussion, we here and later refer to this stage of our method as the \textit{vanilla} VFL-AFE, which effectiveness against privacy attacks is further testified to in~\cref{subsec:protection}. We further make two key remarks: (1) Improving from prior arts, the vanilla VFL-AFE is applicable to generic DNNs as~\cref{eq:norm-clip} solely enforces the range of output embeddings without requiring specific model architecture and loss functions. (2) As a seeming drawback, our derived estimation $\tilde{\Delta}(f)$ is less tight as a trade-off for generality. However, we argue the downgrade can be subsequently reconciled by adaptive adjustments on feature embeddings by their scale and distribution. We concretize our observation in~\cref{subsec:rescale,subsec:distribution}.

\subsection{Adaptive Rescaling}
\label{subsec:rescale}


\begin{figure}[tbp]
  \centering
  \includegraphics[width=\linewidth]{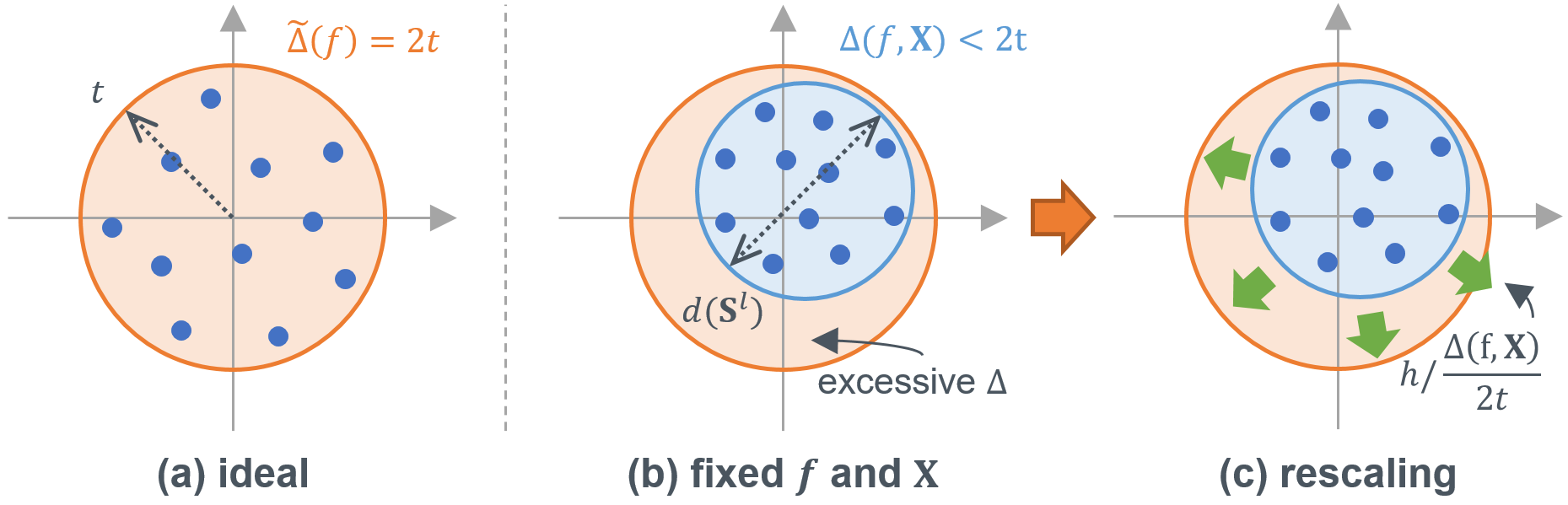}
  \caption{Motivation of rescaling. (a) Ideally, the estimated sensitivity tightly reflects the maximum disparity of embeddings. (b) Excessive sensitivity arises when embeddings follow prior distributions, which produces abundant noise. (c) Rescaling bridges the gap to improve task utility.}
  \Description{The schematic diagram of our proposed rescaling technique.}
  \label{fig:rescaling}
  \vspace{-2mm}
\end{figure}

In this section, we further improve the task utility of VFL-AFE under noise perturbation by rescaling the local feature embeddings. 

Recall high task utility demands tight $\tilde{\Delta}(f)$. We first note $\tilde{\Delta}(f)=2t$ (we omit superscripts $i$ for simplicity) in~\cref{subsec:dp-vfl} presents a conservative, worst-case bound. We illustrate by visualizing the probable range of $f$: Ideally, given an arbitrary $f: \mathbf{D} \rightarrow \mathbb{R}^l$ and any $\mathbf{D}$, the range of $f(\mathbf{D})$ clipped by $t$ is an $l$-dimension ball: $\mathbf{B}^l \triangleq \{ \mathbf{x} \in \mathbb{R}^l : \|\mathbf{x}\| \leq t \}$. Therefore, given neighboring $\mathbf{D},\mathbf{D}'$, the sensitivity $\Delta(f)$ can be concretized as the maximum distance between any two points within $\mathbf{B}^l$ (\textit{i.e.}, its \textit{diameter} $d(\mathbf{B}^l)$), equaling $2\max||h_j||=2t,\forall j$ (\cref{fig:rescaling}(a)). In practice, however,   $f$ is a deterministic function trained from \textit{specific} $\mathbf{X}$ that follows certain prior distributions. The in-uniformity of $f(\mathbf{X})$ would constrain its range to a dense subset $\mathbf{S}^l \subset \mathbf{B}^l$, with believably $d(\mathbf{S}^l)<2t$. (\cref{fig:rescaling}(b)). In other words, there would be ``excessive sensitivity'' from the discrepancy between $\tilde{\Delta}(f)$ and the actual maximum disparity of $f$ regarding specific $\mathbf{X}$, while making up the gap would improve task utility.

We start by characterizing the disparity of $f(\mathbf{X})$ with a relaxed form of $\Delta(f)$. Recall ~\cref{def:sensitivity}  is defined on \textit{arbitrary} $\mathbf{D}$ and is hard to quantify. As we here are curious about the sensitivity regarding  \textit{actual} data $\mathbf{X}$, we leverage the notion of \textit{local sensitivity} as:

\begin{definition}[Local Sensitivity~\cite{DBLP:conf/stoc/NissimRS07}]
\label{def:local-sens}
The local sensitivity of a function $f: \mathbf{X}\rightarrow \mathbb{R}^l$ under fixed $\mathbf{X}$ and any neighbor $\mathbf{X}'$ is:    $$\Delta(f,\mathbf{X})=\max_{\mathbf{X}'} ||f(\mathbf{X})-f(\mathbf{X}')||.$$
\end{definition}

\noindent By~\cref{def:local-sens}, we note: (1) $\Delta(f,\mathbf{X})$ is also applicable for~\cref{lem:gaussian} to allow the calibration of noise~\cite{Dwork14Gaussian}. However, (2) directly replacing $\Delta(f)$ with $\Delta(f,\mathbf{X})$ may lead to potential privacy risks~\cite{Near21ProgrammingDP} as we are to discuss later. 
$\Delta(f,\mathbf{X})$ is more estimable than $\Delta(f)$ as we can associate it with the \textit{diameter} of $f(\mathbf{X})$. Informally, with probability $p_1$ (where $p_1 \rightarrow 1$ when $N$ is large) and a weak assumption on $\mathbf{X}'$(refer to the proof of~\cref{th:vfl-dp-2}), we have:

\begin{equation}
\label{eq:diameter}
\Delta(f, \mathbf{X}) = d(\mathbf{S}^l) \triangleq \max_{j\neq k}||h_j-h_k||.
\end{equation}

\begin{figure}[tbp]
  \centering
  \includegraphics[width=\linewidth]{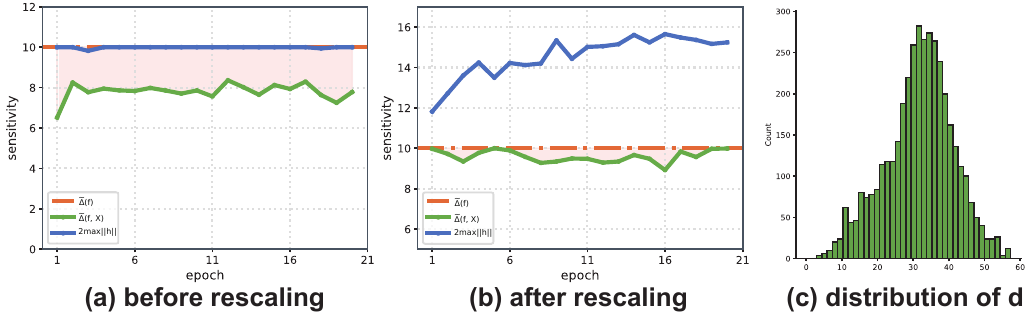}
  \caption{Explanation of rescaling. (a) Before, estimated $\tilde{\Delta}(f)=2t$ (red) bounds $d(\mathbf{B}^l)=2\max||h_j||$ (blue). As typically $\max||h_j-h_k||<2\max||h_j||$, such discrepancy results in excessive sensitivity marked as the red-shaded area. (b) After, $\tilde{\Delta}(f)$ bounds the actual maximum disparity $\max||h_j-h_k||$ (green), making well use of sensitivity. (c) $\mathbb{P}(\mathbf{b})$ empirically follows $\mathcal{N}(\mu_h,\sigma_h^2)$, allowing a statistical estimation of $\tilde{\Delta}(f,\mathbf{X})$.}
  \Description{Analysis of our proposed rescaling technique.}
  \label{fig:rescale-analysis}
  \vspace{-3mm}
\end{figure}

\noindent As $||h_j-h_k||$ is calculable, this allows us to manifest the discrepancy between $\Delta(f, \mathbf{X})$ (representing the ``required'' sensitivity) and $\tilde{\Delta}(f)$ (determining the actual noise scale). As exemplified in~\cref{fig:rescale-analysis}(a), we can observe a quite salient difference between them.

An intuitive approach to minimize $\tilde{\Delta}(f) - \Delta(f, \mathbf{X})$ seems to be replacing $\tilde{\Delta}(f)=2t$ with the calculated $\Delta(f,\mathbf{X})$. However, we note it would face two constraints, regarding privacy and efficiency: (1) As $\Delta(f,\mathbf{X})$ takes specific $\mathbf{X}$ into account, publicly releasing it (by directly calibrating noise from) may reveal information about $\mathbf{X}$. (2) Note $\Delta(f,\mathbf{X})$ varies with $\theta$ (as $\mathbf{h}=f(\mathbf{X},\theta)$). The \textit{de facto} practice to train DNNs is to divide $\mathbf{X}$ into mini-batches and update $\theta$ step-wisely. Therefore, one would have to calculate the diameter of $\mathbf{h}$ regarding the entire $\mathbf{X}$ (as we demand to bound $\mathbf{X}$ as a whole) at each change of $\theta$, which is of prohibitive cost when $N$ is large.

To reconcile privacy, instead of directly adopting $\Delta(f,\mathbf{X})$, we propose to \textit{adaptively rescale local feature embeddings} to $\tilde{\Delta}(f)$ (\cref{fig:rescaling}(c)). Specifically, after performing norm clipping, we calculate $\Delta(f,\mathbf{X})$ by~\cref{eq:diameter} and rescale each feature embeddings as:

\begin{equation}
\label{eq:rescale}
h_j = h_j / \frac{\Delta(f,\mathbf{X})}{2t}, \forall j \in [N].
\end{equation}

\noindent The noise is calibrated from $\tilde{\Delta}(f)$ and added to the rescaled embeddings. \Cref{eq:rescale} allows us to tightly bound the maximum disparity of $\mathbf{h}$ to $\tilde{\Delta}(f)$, which minimizes excessive noise and helps achieve better task utility. \Cref{fig:rescale-analysis}(b) demonstrates the effect after rescaling. Rescaling also mitigates privacy concerns: $\Delta(f,\mathbf{X})$ is never publicly released, and as $\tilde{\Delta}(f)$ is the known, supposed-to-be bound for \textit{any} $\mathbf{D}$ (which includes $\mathbf{X}$) and is consistent with different $\theta$, it reveals very limited information. 

To address efficiency, we propose to approximate a $\tilde{\Delta}(f,\mathbf{X})$ that is easier to calculate and holds with high probability. As a revisit to~\cref{eq:diameter}, $\Delta(f,\mathbf{X})$ represents the maximum of  the \textit{pair-wise distances of feature embeddings} $\mathbf{d}\triangleq\{||h_j-h_k||\}_{j\neq k}$. We consider the probability distribution $\mathbb{P}(\mathbf{d})$ of $\mathbf{d}$ and empirically find $\mathbb{P}(\mathbf{d})$ mostly follows a Gaussian distribution $\mathcal{N}(\mu_h,\sigma_h^2)$, as exemplified in \cref{fig:rescale-analysis}(c). Therefore, we can derive an estimated maximum $\tilde{\Delta}(f,\mathbf{X})$ close to $\Delta(f,\mathbf{X})$ from the cumulative probability of $\mathcal{N}(\mu_h,\sigma_h^2)$. Specifically, we reduce the calculation of $\mathbf{d}$ on the entire $\mathbf{X}$ to that on one of its mini-batch, denoted as  $\mathbf{X}_n\triangleq \{x_j\}_{j=1}^{n}$ (where $\mathbf{X}_n$ is uniformly sampled from $\mathbf{X}$). At each training step, we randomly sample $\mathbf{X}_n$, calculate $\mathbf{d}_n$ regarding its feature embeddings $\mathbf{h}_n$, and estimate $\mu_h,\sigma_h$ from $\mathbb{P}(\mathbf{d}_n)$ through simple statistics. Hence, by the property of Gaussian distribution, an approximated $\tilde{\Delta}_{p_2}(f,\mathbf{X})$ that upper-bounds $||h_j-h_k||$ for arbitrary $j\neq k$ with probability $p_2$ can be calculated by the following quantile function:

\begin{equation}
\label{eq:quantile}
\tilde{\Delta}_{p_2}(f,\mathbf{X})=Q(p_2;\mu_h,\sigma_h) = \mu_h + \sigma_h\sqrt{2}\operatorname{erf}^{-1}(2{p_2}-1),
\end{equation}

\noindent where $\operatorname{erf}^{-1}$ is the inverse error function. As a practical example, choosing $\tilde{\Delta}_{p_2}(f,\mathbf{X})=\mu_h+3\sigma_h$ yields a very confident $p_2\approx 0.9987$. 

We further ensure that $\mathbb{P}(\mathbf{d})$ \textit{is} close to $\mathcal{N}(\mu_h,\sigma_h^2)$ so that the above discussion holds. Specifically, as we aim to minimize the difference between the distributions of $\mathbb{P}(\mathbf{d})$ and $\mathcal{N}(\mu_h,\sigma_h^2)$, we turn it into an optimizable goal regarding their KL divergence $D_{KL}$:

\begin{equation}
\label{eq:kl-divergence}
\mathcal{L}_{KL}(h;\theta) = \alpha \cdot D_{KL}(\mathbb{P}(\mathbf{d})||\mathcal{N}(\mu_h,\sigma_h^2)),
\end{equation}

\noindent where $\alpha$ is the weight, and append it to the VFL objective in~\cref{eq:objective}. We note the actual $\mathcal{L}_{KL}(h,\theta)$ is insignificant among most of our experiments, indicating $\mathbb{P}(\mathbf{d})$ follows $\mathcal{N}(\mu_h,\sigma_h^2)$ quite faithfully.

As a brief summary, this section improves the task utility of VFL-AFE by reducing the discrepancy between the estimated $\tilde{\Delta}(f)$ and the actual $\Delta(f,\mathbf{X})$.  Feature embeddings $\mathbf{h}$ are locally rescaled to mitigate potential privacy leakage and $\tilde{\Delta}(f,\mathbf{X})$ is approximated for efficiency. We reflect \colorbox{blue!20}{our proposed techniques} (\cref{eq:rescale,eq:quantile,eq:kl-divergence}) in~\cref{alg:vfl-afe}. To complete the discussion, we incorporate $p_1,p_2$ into the privacy loss $\delta$, as the exceptional cases they represent could marginally increase the failure probability of DP. We slightly alter our privacy guarantee of VFL (\cref{th:vfl-dp}) as:

\begin{theorem}
\label{th:vfl-dp-2}
The VFL framework specified in~\cref{alg:vfl-afe} and modified by \cref{eq:norm-clip,eq:rescale,eq:quantile,eq:kl-divergence} is $(\epsilon,\delta')$-differentially private, where $\delta'={\delta}/({p_1 p_2})$.
\end{theorem}

\noindent The proof is deferred to supplementary materials.

\subsection{Adaptive Distribution Adjusting}
\label{subsec:distribution}



\begin{figure}[tbp]
  \centering
  \includegraphics[width=\linewidth]{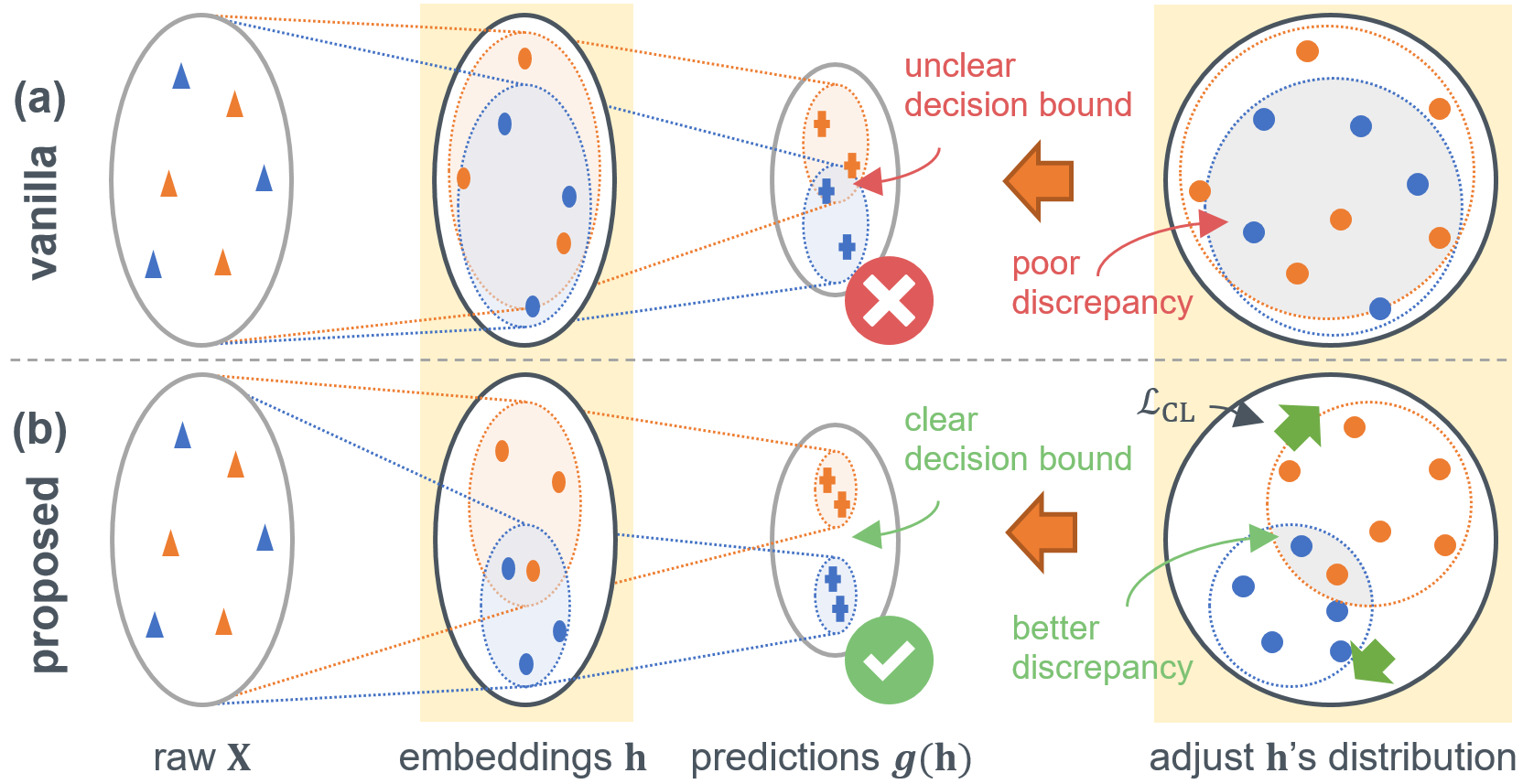}
  \caption{Motivation of distribution adjusting. Enhancing inter-class discrepancy is essential for effective classification. (a) Insufficient discrepancy can lead to ambiguous decision boundaries. (b) By CL, we encourage greater discrepancy among embeddings to improve the final predictions.}
  \Description{The schematic diagram of our proposed adjustments on feature embeddings' distributions.}
  \label{fig:cl}
  \vspace{-2mm}
\end{figure}

To further achieve our goal of improving task utility without compromising the established DP mechanism, we propose to \textit{adaptively adjust the distribution of feature embeddings} in a favorable way for downstream tasks. Specifically, we focus on the classification task, which is the most common scenario in VFL. A classification model typically requires high inter-class and low intra-class discrepancy among its outputs, which is gradually enhanced through hierarchical feature extraction. We note the shared feature embeddings $\mathbf{h}$ can be viewed as the middle output of $g(f(\cdot))$. Therefore, intuitively, it would be beneficial to classification accuracy if $\mathbf{h}$ provides clearer distinguishability among different classes, as shown in~\cref{fig:cl}.

To attain this goal, we employ contrastive learning (CL) to let each passive party adjust the distribution of its $\mathbf{h}$ locally. CL learns useful representations by contrasting similar and dissimilar pairs of samples~\cite{chen2020a,khosla2020supervised}. A significant advantage of CL is its potential to amplify the inter-class discrepancy of learned representations, which aligns with our objectives. Self-supervised CL is commonly used and involves data augmentations on samples to generate pairs, with the reliability of augmentation depending on data types. Instead, we propose a generally applicable, weakly-supervised CL method that considers $\{h_j,h_k\}_{j\neq k}$ from the same/different class(es) as similar/dissimilar pairs.

\begin{figure}[tbp]
  \centering
  \includegraphics[width=0.9\linewidth]{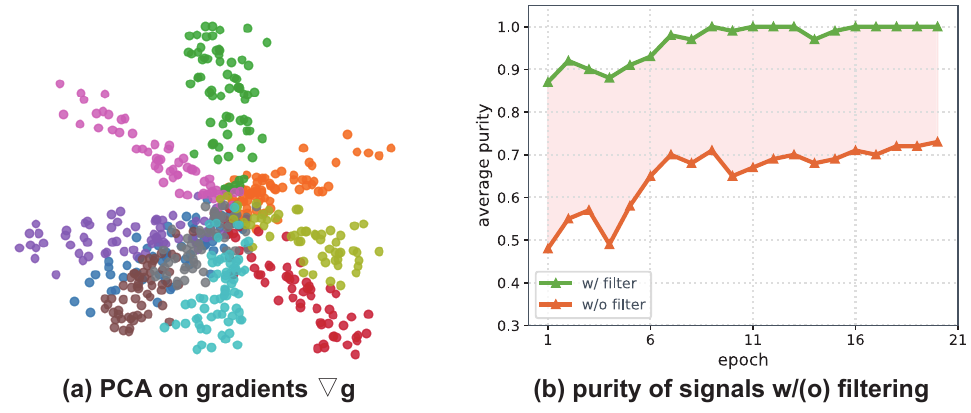}
  \caption{Steps during distribution adjusting. (a) Clustering among exchanged $\nabla g$ can be observed via PCA, enabling the identification of embedding pairs from the same/different class(es) via soft labels. (b) Fuzzy clustering can provide noisy signals regarding ambiguously assigned gradients (red) while filtering helps obtain clearer soft labels (green).}
  \Description{Analysis of our proposed distribution adjusting technique.}
  \label{fig:pca-purity}
  \vspace{-2mm}
\end{figure}

The primary issue to address is to obtain a relatively reliable signal regarding the class, as the labels $\mathbf{y}$ are not accessible for passive parties. To this end, we suggest mining useful information from the active party's exchanged gradients $\nabla g$. Specifically, we argue the magnitude and orientation of $\nabla g$ would imply the sample's class, by the nature of gradient descent. As an illustration, we visualize example gradients regarding a mini-batch via principal component analysis (PCA) in~\cref{fig:pca-purity}(a), where clear clusters can be observed between the gradients of samples from different classes.

By the above observation, we propose to extract \textit{soft labels} for $\mathbf{h}$ from $\nabla g$ leveraging \textit{fuzzy clustering}. In contrast to hard clustering which assigns a specific cluster, fuzzy clustering determines the membership of each sample by a confidence degree within $[0,1]$, which later serves as an effective filter. Specifically, given $\mathbf{h}_n$ of a mini-batch, denote the returned gradients as $\nabla g_n \in \mathbb{R}^{n\times l}$. We presume the passive parties know the number of total classes $C$, which holds in many practical cases. We concretely opt for fuzzy c-means (FCM)~\cite{bezdek1984fcm}, \textit{wlog.}, as our clustering algorithm, and calculate:

\begin{equation}
\label{eq:fcm}
\{\mathbf{I}_n,\mathbf{c}_n\} = \operatorname{FCM}(\nabla g_n, C),
\end{equation}

\noindent where with a slight abuse of notion, $\mathbf{I}_n,\mathbf{c}_n\in \mathbb{R}$ denote the IDs of the most probable membership cluster for $\mathbf{h}_n$ and its confidence, respectively. We note $\mathbf{I}_n\triangleq\{I_j\}_{j=1}^n$ is not corresponding to the true labels $\mathbf{Y}_n$ (as pseudo-labels) since the order of cluster IDs is assigned arbitrarily and changes step-wisely. However, $\mathbf{I}_n$ \textit{does} indicate whether any two $\{h_j,h_k\}$ belong to the same class, which is sufficient to serve as the signal for CL. 

We further note $\nabla g$ provides noisy signals as a portion of gradients could have ambiguous cluster assignments, which compromises the accuracy of $\mathbf{I}_n$. To illustrate, we measure the quality of FCM by \textit{purity}, where high purity indicates correct clustering. In~\cref{fig:pca-purity}(b), the purity considering all $\mathbf{I}_n$ (red line) is not satisfying. To amend, we set up a threshold $c$ to the confidence $\mathbf{c}_n$ to filter ambiguous gradients. We turn $\mathbf{c}_n$ into a 0-1 mask that filters any $\mathbf{I}_n$ with confidence below $c$, and bring in only the rest for CL. Results (\cref{fig:pca-purity}(b)) show the purity of remaining $\mathbf{I}_n$ (green line) is close to 1, indicating a clear signal. 

\begin{figure}[tbp]
  \centering
  \includegraphics[width=0.9\linewidth]{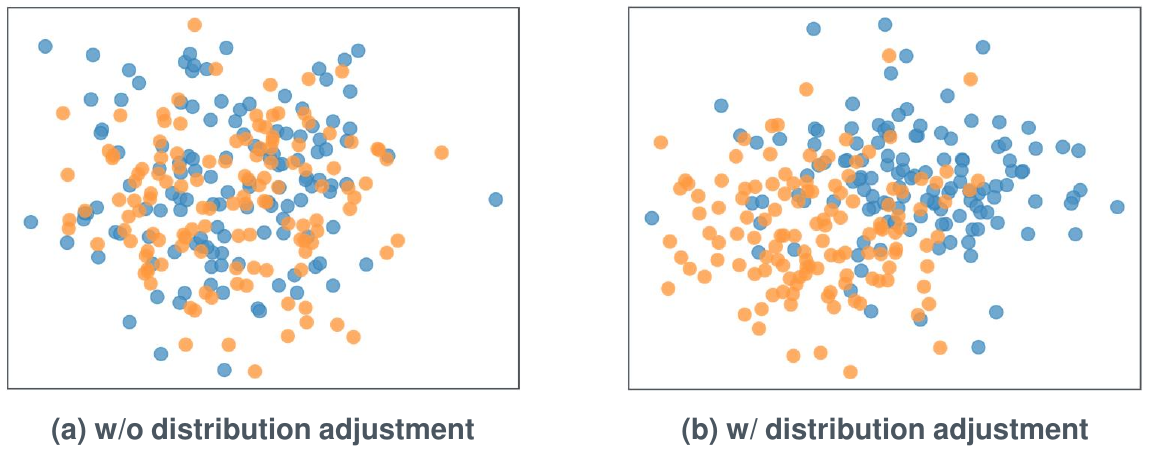}
  \caption{Effect of distribution adjusting via PCA. (a) Without adjusting, feature embeddings from different classes highly overlap regarding distributions. (b) With adjusting, CL encourages better inter-class discrepancy that benefits utility.}
  \Description{Illustration of distribution with and without adjustment via PCA.}
  \label{fig:dist-pca}
  \vspace{-3mm}
\end{figure}

Finally, we perform weakly-supervised CL on feature embeddings $\mathbf{h}_n$ with remaining $\mathbf{I}_n$, to encourage their inter-class discrepancy. For any two $\{h_j,h_k\} \subset \mathbf{h}_n$, let $\omega_{jk}=1$ if $I_j=I_k$ and $\omega_{jk}=0$ otherwise. We establish the CL objective regarding $\mathbf{h}_n$ as:

\begin{equation}
\label{eq:cl}
\mathcal{L}_{CL} = \beta \cdot \frac{1}{n^2} \sum_{j=1}^{n}\sum_{k=1}^{n}(1-\omega_{jk})\left \| h_j - h_k\right \|,
\end{equation}

\noindent where $\beta$ is the weight, and append $\mathcal{L}_{CL}$ to the VFL objective in~\cref{eq:objective}. To illustrate the effect of CL, we exemplify the distribution of two classes of feature embeddings $\mathbf{h}$, before and after appending $\mathcal{L}_{CL}$, via PCA. \Cref{fig:dist-pca} shows CL enhances the discrepancy between the classes, which benefits task utility, as later elaborated in~\cref{subsec:task-utility}. \colorbox{green!20}{Our proposed technique} is reflected in~\cref{alg:vfl-afe}.

We provide additional comments on data privacy: The privacy guarantees of~\cref{th:vfl-dp,th:vfl-dp-2} are also applicable after appending \cref{eq:cl}, due to: (1) all adjustments made to $\mathbf{h}$ are performed privately before the addition of noise, which does not consume privacy budget by the property of DP, and (2) $\tilde{\Delta}(f)$ still bounds $\mathbf{h}$ as it only concerns the maximum disparity of $\mathbf{h}$ (which remains the same) and does not consider the distribution within the bound. This echoes our purpose, to improve task utility in a flexible way, with no/minimal change(s) to established DP mechanisms.



\section{Experiments}
\label{sec:experiments}

In this section, we demonstrate VFL-AFE can be generally applied to different datasets and model structures, while the adaptive feature embeddings effectively improve task utility. In addition to formal DP guarantees (\cref{th:vfl-dp,th:vfl-dp-2}), we experimentally demonstrate that VFE-AFE is resilient against common privacy threats.

\subsection{Experimental Setups}
\label{subsec:setups}

\begin{figure}[tbp]
  \centering
  \includegraphics[width=\linewidth]{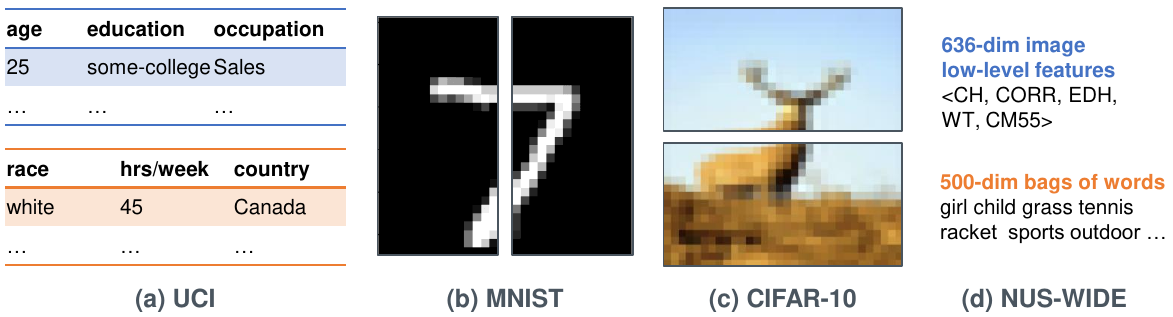}
  \caption{Example training data and their partitions.}
  \Description{Example data for our experimental setups.}
  \label{fig:example-data}  
\vspace{-4mm}
\end{figure}

\noindent \textbf{Datasets.} To elaborate on the generality of VFL-AFE, we employ 4 datasets that contain structural, image, and textual data, specifically: (1) \textbf{UCI}~\cite{DuaG17UCI}, the Adult dataset that contains records on 48K individuals, with attributes such as age, education level, and occupation, and a binary label regarding income. We split records by attributes to simulate VFL among multi-sources. (2) \textbf{MNIST}~\cite{Deng2012mnist}, the handwriting digits dataset that contains 60K gray-scale images from 10 classes. (3) \textbf{CIFAR-10}~\cite{krizhevsky2009cifar}, which contains 60K colored images from 10 classes of objects. We cut the images into vertical and horizontal halves for MNIST and CIFAR-10, respectively, to simulate multi-views. (4) \textbf{NUS-WIDE}~\cite{ChuaTHLLZ09NUSWide}, a multi-modality dataset with image and textual features on 270K Flickr images divided into 81 concepts, where we leverage the top 10 concepts. We partition the data by modality. \Cref{fig:example-data} exemplifies the data and their partitions.

\noindent \textbf{Models architectures.} For UCI and NUS-WIDE, we employ multiple linear layers as $f$ and a linear regression head as $g$. For MNIST and CIFAR-10, we employ a two-layer convolution neural network (CNN) and LeNet-5~\cite{Lecun98LeNet} with feature flattening as $f$, respectively. We take a linear layer plus softmax head as $g$.

\noindent \textbf{Implementation details.} We consider the collaboration, \textit{wlog.}, between the active party and 2 passive parties. Each passive party calculates its noise scale and applies the DP mechanism independently. We fix privacy loss $\delta$=1e-2. We apply relatively generous budgets $\epsilon$ to ensure accuracy, whereas experimental results show VFL-AFE provides effective resiliency against SOTA privacy attacks under our choice of $(\epsilon,\delta)$. We choose $\lambda$=1e-4, and $\alpha,\beta$ that align with the order of magnitude of the primary VFL objective. The same random seed is sampled across all experiments.


\subsection{Task Utility}
\label{subsec:task-utility}

\begin{figure}[tbp]
  \centering
  \includegraphics[width=0.9\linewidth]{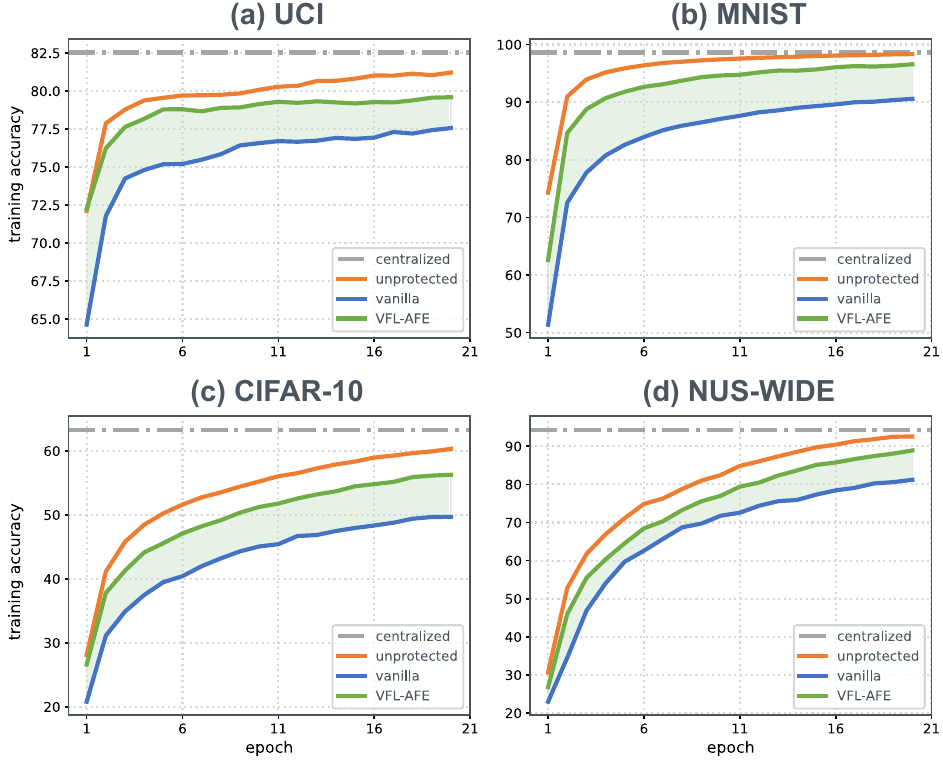}
  \caption{Performance of VFL-AFE. Vanilla method (blue) enhances privacy by noise distribution, which inevitably reduces model accuracy compared to unprotected baselines (red). We address the downgrade flexibly via rescaling and distribution adjusting, which enhance task utility (green). The green-shaded region marks the significant accuracy gain. Similar trends can be observed among all datasets, testifying to the generality of our method.} 
  \Description{Task utility of our proposed method.}
  \label{fig:task-utility}
  \vspace{-6.5mm}
\end{figure}

We train 4 models with respect to each dataset: (1) a centralized baseline, (2) an unprotected VFL  (\textit{i.e.}, VFL without DP), (3) the vanilla VFL-AFE in~\cref{subsec:dp-vfl} (\textit{i.e.}, VFL with DP yet without adaptive feature embeddings), and (4) our proposed VFL-AFE. Each model is trained for 20 epochs with the learning rate $lr$=1e-3, 1e-5, 1e-4, 1e-4 for UCI, MNIST, CIFAR-10, and NUS-WIDE, respectively. \Cref{fig:task-utility} shows the epoch-wise training accuracy, where we note: (1) Unprotected VFL achieves close accuracy to the centralized model. It yet fails to defend against privacy attacks, as later shown in~\cref{subsec:protection}. (2) The vanilla VFL-AFE experiences a significant accuracy drop among 4 \textasciitilde 11\%, owing to the noise perturbation of DP mechanisms. (3) Nonetheless, we demonstrate the utility loss can be largely mitigated by our adaptive adjustments on feature embeddings, as the accuracy of the final VFL-AFE significantly improves by 2 \textasciitilde 7\% compared to vanilla. This suggests VFL-AFE aligns with our goal, \textit{i.e.}, to address the privacy-utility balance in a more flexible way.


\subsection{Data Privacy Against Threats}
\label{subsec:protection}

\begin{figure}[tbp]
  \centering
  \includegraphics[width=0.95\linewidth]{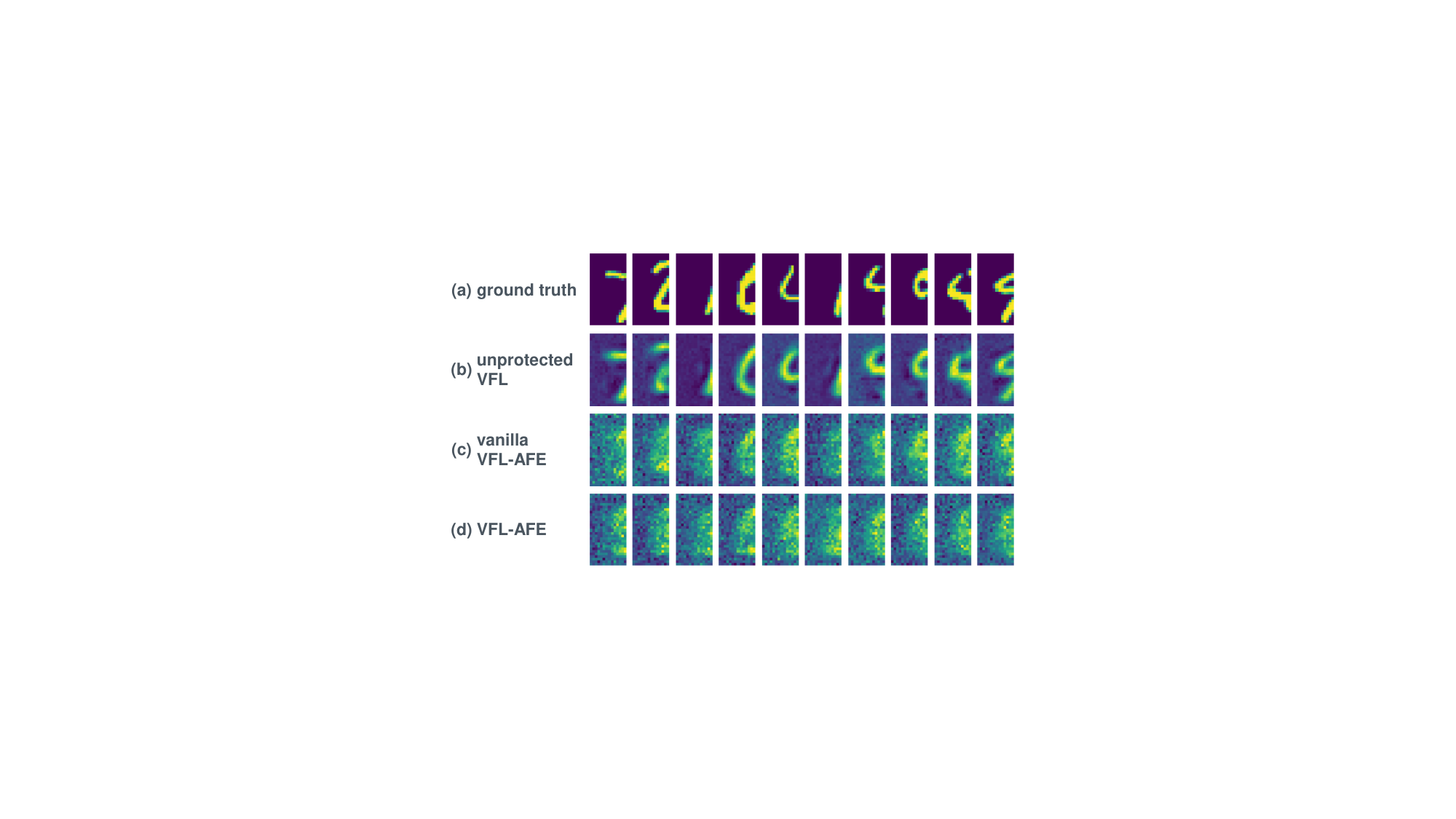}
  \caption{Resiliency against inversion attacks. (b) Unprotected VFL provides deficient defense, as the recovered images are close to (a) ground truth. (c-d) Both vanilla and final VFL-AFE defend the attack effectively. Notably, a similar level of protection is retained after utility-improving measures.}
  \Description{The effectiveness of our proposed method against inversion attacks.}
  \label{fig:inversion}
  \vspace{-2mm}
\end{figure}

In addition to theoretical analyses, we experimentally study the privacy protection capability of VFL-AFE. The purpose of DP is to protect data confidentiality against privacy threats, namely, inversion and membership inference (MI) attacks. We here compare the unprotected, vanilla, and final VFL-AFEs under SOTA attacks~\cite{DBLP:conf/sp/ShokriSSS17,DBLP:journals/corr/YeJWLL22recon}.

\noindent \textbf{Inversion attack.} The attacker aims to recover original samples $\mathbf{X}$ of a victim passive party from the shared embeddings $\mathbf{h}$~\cite{DBLP:journals/corr/YeJWLL22recon}. We assume the attacker possesses some samples $\mathbf{X}_{atk}$ that share similar distribution with $\mathbf{X}$ and can query the victim's $f$ infinitely. Hence, it can train a decoder $f^{-1}$ by minimizing $|| f^{-1}(f(\mathbf{X}_{atk}))-\mathbf{X}_{atk}||$, and exploit the trained model on any received $\mathbf{h}$. We analyze the attack on MNIST: For each trained VFL model, we let train such an $f^{-1}$ till it converges. By the visualization of results in~\cref{fig:inversion}, we note: (1) Unprotected VFL shows almost no resiliency to inversions. (2) Both vanilla and final VFL-AFE profoundly safeguard $\mathbf{X}$ from being revealed, as the recovered images are highly blurred. (3) Notably, they show similar capabilities in protection. This supports that our adaptive feature embedding techniques do not affect the protection of established privacy protections.

\begin{table}[tbp]
  \caption{Attacker's accuracy of membership inference.} 
  \centering
  \begin{tabular}{@{}lcccc@{}}
    \toprule
    \multirow{2}{*}{\textbf{Method}} & \multicolumn{4}{c}{\textbf{MI Attack Accuracy} $\boldsymbol{\downarrow}$} \\
    \cmidrule(lr){2-5}
     & \multicolumn{1}{p{1.2cm}}{\centering \textbf{UCI}} & \multicolumn{1}{p{1.2cm}}{\centering \textbf{MNIST}} & \multicolumn{1}{p{1.2cm}}{\centering \textbf{CIFAR}} & \multicolumn{1}{p{1.2cm}}{\centering \textbf{NUS-W}} \\
    \midrule
    \textbf{unprotected} & 51.52 & 62.36 & 65.75 & 71.19 \\
    \textbf{vanilla} & 51.03 & 51.75 & 52.32 & 54.70 \\
    \textbf{VFL-AFE} & 51.12 & 53.19 & 54.65 & 55.23 \\
    \bottomrule
  \end{tabular}
  \label{tab:mia}
\vspace{-3mm}
\end{table}

\noindent \textbf{Membership inference attack.} The attacker aims to determine whether a specific $x$ belongs to the training data $\mathbf{X}$. To this end, ~\cite{DBLP:conf/sp/ShokriSSS17}  proposes a two-step attack, to train multiple \textit{shadow models} that mimic the behavior of the victim's $f$, and an \textit{attack model} that speculates the membership. We use an open-source implementation of the attack~\cite{AIJack} and report the attacker's accuracy (lower-bounded by 50\%) on VFLs in~\cref{tab:mia}. Lower accuracy indicates better resiliency. The attack is effective on unprotected VFL for all datasets except UCI due to limited exploitable information of binary labels (further see~\cite{DBLP:conf/sp/ShokriSSS17}). We remark: (1) Both vanilla and final VFL-AFE provide effective defenses against the attacks, as the attack accuracy is reduced to close to 50\%. This testifies to the privacy protection of our DP mechanism. (2) The accuracy is slightly higher in the final VFL-AFE. We speculate it as a balance for accuracy, as the increased inter-class discrepancy (\cref{subsec:distribution}) is also favorable for the shadow models. Nonetheless, we note the trade-off is marginal and our DP guarantees still hold.


\subsection{Ablation Study}
\label{subsec:abal}

\begin{table}[tbp]
  \caption{Contribution of each component to accuracy.}
  \centering
\begin{tabular}{@{}lcccc@{}}
    \toprule
    \multirow{2}{*}{\textbf{Method}} & \multicolumn{4}{c}{\textbf{Test Accuracy}}\\
    \cmidrule(lr){2-5}
     & \multicolumn{1}{p{1.2cm}}{\centering \textbf{UCI}} & \multicolumn{1}{p{1.2cm}}{\centering \textbf{MNIST}} & \multicolumn{1}{p{1.2cm}}{\centering \textbf{CIFAR}} & \multicolumn{1}{p{1.2cm}}{\centering \textbf{NUS-W}} \\
    \midrule
\textbf{vanilla}                                     & 77.16        & 90.48          & 48.83             & 66.87             \\
\textbf{vanilla+R}                                   & 79.02        & 95.12          & 54.58             & 70.63             \\
\textbf{vanilla+D}                                   & 77.70        & 92.51          & 49.53             & 68.01             \\
\textbf{VFL-AFE}                                     & \textbf{79.31}        & \textbf{96.53}          & \textbf{55.06}             & \textbf{71.18}             \\
\bottomrule
\end{tabular}
  \label{tab:abal}
  \vspace{-3mm}
\end{table}

We analyze the independent contribution of rescaling (\cref{subsec:rescale}) and distribution adjusting (\cref{subsec:distribution}) to task utility. Results are summarized in~\cref{tab:abal} by test accuracy, where ``vanilla+R'' and ``vanilla+D'' represents the results of rescaling and distribution adjusting alone, respectively.  We note: (1) Rescaling contributes the majority of utility gain. However, applying it is at the cost of higher run-time overheads (\cref{subsec:time-cost}). Distribution adjusting also demonstrates stable effects with relatively low computational costs. (2) Both proposed techniques can be employed together to achieve better task utility. 

We defer further ablation studies to supplemental materials due to space limits, where we assume the readers may be interested in some key information, \textit{e.g.}, the choice of thresholds $t,c$.

\subsection{Computation Overheads}
\label{subsec:time-cost}

\begin{table}[tbp]
\caption{Computational cost by training time.}
\begin{tabular}{lcccc}
\toprule
\textbf{}          & \textbf{VFL} & \textbf{+noise} & \textbf{+R} & \textbf{+D} \\
\midrule
\textbf{time (ms)} & 349.89       & 60.43       & 279.76     & 82.04      \\
\textbf{time (\%)} & 45.32\%      & 7.83\%      & 36.23\%    & 10.63\% \\
\bottomrule
\end{tabular}
\label{tab:time-cost}
\vspace{-3mm}
\end{table}

\Cref{tab:time-cost} demonstrate the time cost of 5000 CIFAR-10 samples for baseline VFL, the addition of noise, rescaling (+R), and distribution adjusting (+D), respectively. We observe that the computation of pair-wise distances in rescaling consumes the most considerable time, which, nonetheless, could be believably improved by optimizing algorithms. Overall, we argue the time cost is within a decent scope regarding improved task utility. It is further worth noting that our method requires no extra communication rounds and overheads, which is more favorable than some prior arts~\cite{10.1145/3467956,DBLP:journals/corr/HardyHINPST17VFLHE,DBLP:journals/corr/YangFCSY19QuasiNewton,DBLP:conf/ccs/XuB00JL21}.



\section{Conclusion}
\label{sec:conclusion}

This paper studies the trade-off between data privacy and task utility in VFL under DP. In the proposed VFL-AFE framework, we initially derive a rigorous and generic DP privacy guarantee by performing norm clipping on shared feature embeddings. We subsequently reconcile its utility downgrade by the proposed adaptive feature embeddings, where rescaling and distribution adjusting are conducted to benefit model performance. Our takeaway message is: DP and VFL can be combined in a more flexible way.

\bibliographystyle{ACM-Reference-Format}
\bibliography{mm23}

\end{document}